\documentclass[letterpaper]{article} % DO NOT CHANGE THIS
\usepackage{aaai2026}  % DO NOT CHANGE THIS
\usepackage{times}  % DO NOT CHANGE THIS
\usepackage{helvet}  % DO NOT CHANGE THIS
\usepackage{courier}  % DO NOT CHANGE THIS
\usepackage[hyphens]{url}  % DO NOT CHANGE THIS
\usepackage{graphicx} % DO NOT CHANGE THIS
\usepackage{natbib}  % DO NOT CHANGE THIS AND DO NOT ADD ANY OPTIONS TO IT
\usepackage{caption} % DO NOT CHANGE THIS AND DO NOT ADD ANY OPTIONS TO IT
\usepackage{algorithm}
\usepackage{algorithmic}
\usepackage{multirow}
\usepackage{threeparttable} 
\usepackage{booktabs}
\usepackage{amssymb}
\usepackage{amsmath}
\usepackage{graphicx}
\usepackage{color}

\usepackage{newfloat}
\usepackage{listings}
\DeclareCaptionStyle{ruled}{labelfont=normalfont,labelsep=colon,strut=off} % DO NOT CHANGE THIS
\floatstyle{ruled}
\newfloat{listing}{tb}{lst}{}
\floatname{listing}{Listing}
\title{SACodec: Asymmetric Quantization with Semantic Anchoring for Low-Bitrate High-Fidelity Neural Speech Codecs}
\author{
    Zhongren~Dong\textsuperscript{\rm 1}, Bin~Wang\textsuperscript{\rm 2}, Jing~Han\textsuperscript{\rm 1}\thanks{Corresponding author.}, Haotian~Guo\textsuperscript{\rm 1}, Xiaojun~Mo\textsuperscript{\rm 1}, Yimin~Cao\textsuperscript{\rm 1}, Zixing~Zhang\textsuperscript{\rm 1,\rm 3}\\
}
\affiliations{
    \textsuperscript{\rm 1}College of Computer Science and Electronic Engineering, Hunan University, Changsha, China\\
    \textsuperscript{\rm 2}Beijing Xiaomi Mobile Software Co., Ltd, Beijing, China\\
    \textsuperscript{\rm 3}Shenzhen Research Institute, Hunan University, Shenzhen, China\\
    jhan@hnu.edu.cn
    
}

\usepackage{bibentry}
\begin{document}

\maketitle

\begin{abstract}
Neural Speech Codecs face a fundamental trade-off at low bitrates: preserving acoustic fidelity often compromises semantic richness. To address this, we introduce SACodec, a novel codec built upon an asymmetric dual-quantizer that employs our proposed \textbf{S}emantic \textbf{A}nchoring mechanism. This design strategically decouples the quantization of \textbf{S}emantic and \textbf{A}coustic details. The semantic anchoring is achieved via a lightweight projector that aligns acoustic features with a frozen, large-scale mHuBERT codebook, injecting linguistic priors while guaranteeing full codebook utilization. Sequentially, for acoustic details, a residual activation module with SimVQ enables a single-layer quantizer (acoustic path) to faithfully recover fine-grained information. At just 1.5\,kbps, SACodec establishes a new state of the art by excelling in both fidelity and semantics: subjective listening tests confirm that its reconstruction quality is perceptually highly comparable to ground-truth audio,
while its tokens demonstrate substantially improved semantic richness in downstream tasks. 
\end{abstract}

\begin{links}
    \link{Code}{https://github.com/SmileHnu/SACodec}
\end{links}

\section{Introduction}

The deepening integration of Large Language Models (LLMs) into the speech domain has made Neural Speech Codecs (NSCs)—the crucial bridge between continuous waveforms and discrete tokens—increasingly pivotal~\cite{moshi, qwen2.5omni, kimi-audio}. By converting high-dimensional signals into low-dimensional symbol sequences, discrete tokens form the bedrock of modern Speech Language Models (SLMs), enabling powerful LLM architectures to be applied directly to tasks like Text-to-Speech (TTS) synthesis~\cite{valle, valle2, cosyvoice3} and spoken dialogue~\cite{audiolm, speechgpt, ma2025language}. Consequently, developing low-bitrate codecs has emerged as a central research frontier. 
Traditionally, low bitrates played a role in reducing communication/storage costs; in the LLM era, their value has expanded to enhancing computational efficiency in large-scale models. Efficient codecs mitigate the quadratic complexity of attention mechanisms by generating shorter token sequences, thereby reducing inference latency and cost, which is crucial for real-time audio-language model services~\cite{single-codec, bigcodec}.

Yet, the pursuit of ever-lower bitrates exposes a fundamental trade-off in high-fidelity codecs like Encodec~\cite{encodec} and DAC~\cite{dac}, which rely on multi-layer Residual Vector Quantization (RVQ)~\cite{soundstream}. 
While effective across a range of bitrates, RVQ's performance degrades sharply when the bitrate budget is constrained to 1.5\,kbps. The accumulation of quantization errors across layers yields audible artifacts~\cite{guo2025recent}, and more critically, the resulting multi-stream tokens introduce significant downstream modeling complexity. SLMs now employ intricate parallel or non-autoregressive decoders to handle these hierarchical token streams~\cite{valle, audiolm}, a stark mismatch with the community's drive for the architectural simplicity of a single, unified sequence.

To mitigate this modeling complexity, recent works led by WavTokenizer~\cite{wavtokenizer} have pioneered the single-codebook paradigm. 
By compressing all information into a single discrete sequence, these codecs greatly simplify downstream integration, enabling a more direct application of standard language modeling.
However, this acoustics-oriented optimization, driven solely by signal-distortion objectives, yields representations that lack explicit semantic structure. This becomes a liability for tasks requiring deep content understanding, shifting the challenge from ``how to compress efficiently'' to the more profound question of ``how to encode meaningful content''~\cite{speechtokenizer}.

To infuse codecs with semantics, the community has explored two dominant strategies: {external knowledge distillation}~\cite{speechtokenizer, moshi} and {endogenous self-supervised learning}~\cite{unicodec}. Both routes, however, lead to an uneasy compromise. To preserve reconstruction quality, these methods either retain a complex multi-layer RVQ backend, conflicting with the low-bitrate objective, or introduce significant training complexity and computational overhead. 
Fundamentally, to endow tokens with semantic capacity, the selected codebooks from the quantizer must carry semantic information, rather than serving merely for audio reconstruction. However, traditional VQ is inherently inefficient due to its local update rule—only the nearest codebook is updated—leading to codebook collapse and severely limiting both the expressivity and scalability of the learned codebook~\cite{vqgan, vqgan-lc, simvq}. Breaking this stalemate thus demands a new quantization paradigm. Recent advancements like SimVQ~\cite{simvq}, which reparameterize the codebook to enable global updates, have shown a promising path toward achieving high codebook utilization. A key challenge, which we address, is how to seamlessly couple such efficient quantization with a direct mechanism for semantic injection.

To overcome the twin bottlenecks of VQ inefficiency and cumbersome semantic injection, we introduce a novel \textbf{S}emantic-\textbf{A}nchored speech codec, namely \textbf{SACodec}. Our approach is founded on a \textit{asymmetric dual-quantizer} architecture that assigns specialized, highly efficient quantization mechanisms to distinct semantic and acoustic information streams. By strategically decoupling the modeling of this information at the quantization level, we address the core trade-offs that constrain existing codecs. Our contributions are threefold:

\begin{itemize}
    \item We propose a novel semantic anchoring mechanism that leverages a fixed, large-scale mHuBERT codebook. A lightweight learned projector adaptively aligns this external knowledge to the acoustic latent space, efficiently injecting strong semantic priors while effectively preventing codebook collapse in the semantic layer.
    \item We introduce a synergistic residual activation module that equips a single-layer VQ with the SimVQ technique. This design guarantees full codebook activation for the residual quantizer, enabling it to compensate for fine-grained acoustic details with minimal architectural complexity and bitrate overhead.
    \item  We demonstrate through extensive experiments that SACodec establishes a new record for low-bitrate speech codecs. At 1.5\,kbps, it achieves superior reconstruction quality over all baselines and exhibits stronger semantic representation in downstream tasks, offering a better-balanced solution for modern SLMs.
\end{itemize}

\section{Related Works}
Recent advances in NSCs have evolved along three main dimensions: acoustic codec paradigms, semantic enhancement strategies, and training paradigms. We review each to contextualize SACodec's contributions.

\textbf{Acoustic Codec Paradigms.} Dominant high-fidelity NSCs (SoundStream~\cite{soundstream}, Encodec~\cite{encodec}, DAC~\cite{dac}, HiFi-Codec~\cite{hificodec}) rely on RVQ. By cascading multiple codebooks, RVQ-based models excel at reconstruction but produce multi-stream tokens, which complicates downstream autoregressive modeling~\cite{guo2025recent}.
% To address this, a recent trend has moved towards single-codebook codecs, exemplified by WavTokenizer~\cite{wavtokenizer}, BigCodec~\cite{bigcodec}, and Single-Codec~\cite{single-codec}. These models output a single token sequence, greatly simplifying integration with language models and achieving remarkable compression ratios.
Recent single-codebook codecs (WavTokenizer~\cite{wavtokenizer}, BigCodec~\cite{bigcodec}, Single-Codec~\cite{single-codec}) output single token sequences, simplifying integration with language models and enabling efficient compression.
However, a fundamental challenge shared by both multi-layer and single-layer approaches is codebook collapse, where only a fraction of the learnable codebook is utilized, capping their ultimate representational power~\cite{vqgan, vqgan-lc, simvq}.

\textbf{Semantic Enhancement Strategies.} To address the semantic sparsity of acoustic-only codecs, two main strategies have emerged. 
The first relies on external knowledge distillation. Models like SpeechTokenizer~\cite{speechtokenizer} and Mimi~\cite{moshi} use pre-trained SSL models (e.g., HuBERT~\cite{hubert}, WavLM~\cite{wavlm}) as “teachers” to guide early quantization. This reflects a broader trend of injecting semantic priors, with some works even employing fixed semantic codebooks from text models like LLaMA~\cite{llm-codec}.
While effective, this strategy introduces a significant dependency on external, often large-scale, pre-trained models, and still requires a multi-layer RVQ backend to compensate for the potential degradation in reconstruction quality, which conflicts with low-bitrate goals.

The second route pursues endogenous semantic learning. This includes methods ranging from disentangling speech attributes via multi-task supervision in FACodec~\cite{facodec} to decoupling speaker timbre in LSCodec~\cite{lscodec}. A parallel line of work also explores diffusion-based models like SemantiCodec~\cite{semanticodec}, which generate acoustics from semantic tokens. These approaches, while more self-contained, often lead to a different set of trade-offs: the disentanglement process can be fragile and hard to optimize, while diffusion-based decoders introduce substantial computational overhead during inference, limiting their applicability in real-time scenarios.

\textbf{Training Paradigms and the Unifying Dilemma.} From a training perspective, the dual objectives of acoustic fidelity and semantic richness have led to complex paradigms. Many advanced codecs adopt staged or intricate multi-task frameworks to balance these competing goals. For instance, some approaches rely on multi-stage pipelines~\cite{semanticodec}, while others require separate modules for semantic and acoustic modeling~\cite{facodec}.
Although effective, such non-monolithic designs often hinder reproducibility and complicate optimization, as highlighted in recent benchmarks~\cite{mousavi2025discrete}. 
Conversely, while simple end-to-end paradigms are attractive, they have traditionally struggled to integrate semantic information effectively. Even in tasks like TTS, bridging the gap between textual and acoustic tokens frequently requires complex cascaded models~\cite{spear-tts}.

These challenges stem from a common root: the inefficiency of conventional VQ. This forces a compromise among architectural simplicity, direct semantic injection, and low-bitrate performance. SACodec confronts this unifying bottleneck directly. By integrating two distinct, highly efficient quantization mechanisms—one for anchoring semantics and one for activating residuals—within a single end-to-end framework, our work offers a novel approach to this long-standing trilemma.

\begin{figure*}[t]
\centering
\includegraphics[width=1\textwidth]{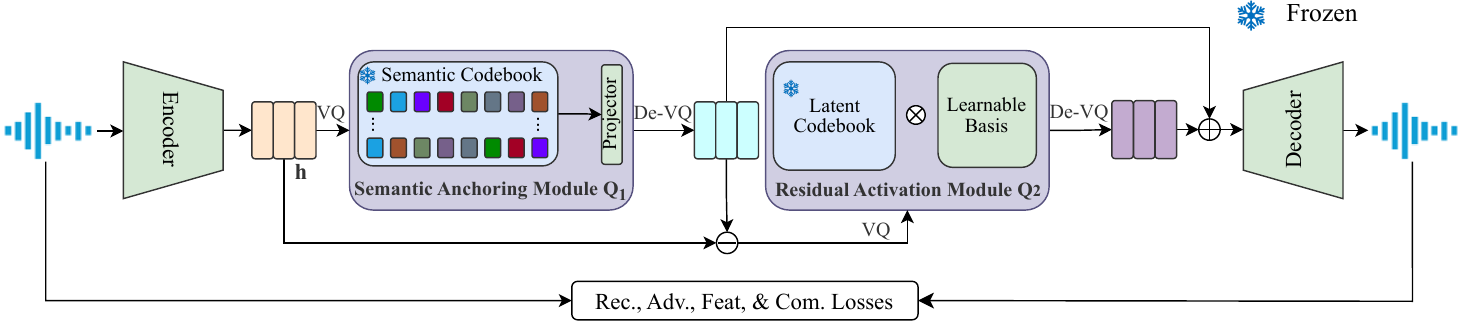} 
\caption{The architecture of SACodec, centered on our \textbf{Asymmetric Dual Quantizer}. An input waveform is mapped by a convolutional-recurrent \textbf{Encoder} to a latent representation $\mathbf{h}$. This representation is processed sequentially by two specialized quantization modules: 
\textbf{(1) Semantic Anchoring (Q1):} A learnable Projector aligns $\mathbf{h}$ with a large-scale, pre-computed, and frozen Semantic Codebook derived from mHuBERT. This process anchors the core linguistic content. 
\textbf{(2) Residual Activation (Q2):} The acoustic residual ($\mathbf{h}$ minus the semantic embedding) is quantized by a single-layer VQ. This quantizer employs the SimVQ technique, where a Learnable Basis transforms a randomly initialized and frozen Latent Codebook to dynamically form the residual space, ensuring full codebook activation. 
During training, only the Projector and the Learnable Basis are updated. The outputs of both quantizers are summed and passed to ConvNeXt-Attention \textbf{Decoder}, which reconstructs the speech signal via an iSTFT.}
\label{fig:framework}
\end{figure*}

\section{Methodology}
% We introduce SACodec, a novel neural speech codec designed for low-bitrate, high-fidelity, and semantically rich speech tokenization. This section first presents the overall architecture, detailing the encoder, decoder, and our core innovation: an Asymmetric Dual Quantizer. We then dissect the two key components of this quantizer: a Semantic Anchoring module for semantic representation and a Residual Activation module for acoustic detail compensation. Finally, we formulate the joint training objective that unifies these components in an end-to-end framework.
This section details SACodec, a novel neural speech codec for low-bitrate, high-fidelity, and semantically rich tokenization. We first present the overall architecture, then dissect its core innovation—the Asymmetric Dual Quantizer—and finally formulate the unified training objective.

% \vspace{-.2cm}
% \subsection{Overall Framework}
% SACodec is built upon a GAN-based~\cite{gan}, end-to-end framework, with its full pipeline illustrated in Fig.~\ref{fig:framework}. The model comprises three core components: an encoder, our proposed asymmetric dual quantizer, and a decoder.

% Given an input speech waveform $\mathbf{x} \in \mathbb{R}^L$, the encoder first downsamples it into a continuous latent representation $\mathbf{h} \in \mathbb{R}^{T \times D}$. Subsequently, this representation is processed by our asymmetric dual quantizer in two sequential stages. First, the Semantic Anchoring Module ($\mathbf{Q}_1$) quantizes $\mathbf{h}$ against a projected, frozen mHuBERT codebook to extract the core semantic content, yielding an embedding $\mathbf{e}_1$. The resulting acoustic residual, $\mathbf{r} = \mathbf{h} - \mathbf{e}_1$, is then quantized by the Residual Activation Module ($\mathbf{Q}_2$), which uses the SimVQ technique to efficiently capture fine-grained acoustic details in a second embedding, $\mathbf{e}_2$.
% 

Finally, the quantized embeddings from both modules, $\mathbf{e}_1$ and $\mathbf{e}_2$, are fused by element-wise addition to form the final representation $\mathbf{e}_{\text{final}} = \mathbf{e}_1 + \mathbf{e}_2$. This is passed to the decoder to reconstruct the high-fidelity waveform $\hat{\mathbf{x}}$. The entire generator is trained adversarially against an ensemble of multi-scale and multi-period discriminators.

% \subsection{Encoder}
% Our encoder, following the Encodec backbone~\cite{encodec}, employs a convolutional stack with Snake activations~\cite{dac} and a two-layer LSTM~\cite{lstm}. It processes a 24\,kHz waveform through a series of strided convolutions, achieving a 320$\times$ downsampling to a 75\,Hz frame rate. A final linear layer then projects the features to the target dimension $D$, producing the latent representation $\mathbf{h}$.

\subsection{Overall Framework}
SACodec is built upon a GAN-based~\cite{gan}, end-to-end framework, with its full pipeline illustrated in Fig.~\ref{fig:framework}. The model comprises three core components: an encoder, our proposed asymmetric dual quantizer, and a decoder, which operate sequentially.

First, the \textbf{encoder} processes an input speech waveform $\mathbf{x} \in \mathbb{R}^L$. Following the Encodec backbone~\cite{encodec}, it employs a convolutional stack with ELU activations and a two-layer LSTM~\cite{lstm}. The encoder processes a 24\,kHz waveform through a series of strided convolutions, achieving a 320$\times$ downsampling to a 75\,Hz frame rate. A final linear layer then projects the features to the target dimension $D$, producing the continuous latent representation $\mathbf{h} \in \mathbb{R}^{T \times D}$.

Next, this latent representation $\mathbf{h}$ is processed by our \textbf{asymmetric dual quantizer}. It operates in two stages: the Semantic Anchoring Module $\mathbf{Q}_1$  first extracts the core semantic content by quantizing $\mathbf{h}$ against a projected mHuBERT codebook, yielding an embedding $\mathbf{e}_1$. The resulting acoustic residual is then quantized by the Residual Activation Module $\mathbf{Q}_2$, which uses the SimVQ technique to capture fine-grained acoustic details in a second embedding, $\mathbf{e}_2$.

Finally, the quantized embeddings from both modules are fused by element-wise addition ($\mathbf{e}_{\text{final}} = \mathbf{e}_1 + \mathbf{e}_2$) and passed to the \textbf{decoder}. Our decoder is engineered for high-fidelity synthesis from this fused representation. Inspired by modern vocoders~\cite{wavtokenizer, vocos}, it decouples feature processing from signal synthesis. A powerful ConvNeXt-Attention backbone first models both local and global dependencies in the feature sequence. The resulting features are then projected to a complex spectrogram and deterministically converted to the output waveform $\hat{\mathbf{x}}$ via an inverse Short-Time Fourier Transform (iSTFT).

The entire generator (encoder, quantizer, and decoder) is trained adversarially against an ensemble of multi-scale and multi-period discriminators.
% All architectural details are provided in Appendix A.

\subsection{Asymmetric Dual Quantizer}
\paragraph{Semantic Anchoring Module}
To overcome the codebook collapse that plagues traditional learnable VQ~\cite{vqgan-lc, simvq} and to directly inject strong semantic priors, our semantic quantizer ($\mathbf{Q}_1$) is built upon a fixed external knowledge base. Note, we adopt the publicly available semantic codebook $\mathbf{C}_{\text{sem}} \in \mathbb{R}^{K_1 \times D_s}$, which consists of $K_1$=1000 centroids clustered from mHuBERT features~\cite{textless}, serving as a stable ``semantic anchor'' for our model.

To bridge the distributional gap between the encoder's acoustic representation $\mathbf{h}$ and the fixed semantic space, we adopt a codebook-space projection strategy~\cite{vqgan-lc}. We learn a lightweight linear projector $\mathbf{P}_{\text{sem}}$ that transforms the \textit{entire} frozen codebook $\mathbf{C}_{\text{sem}}$ into a dynamically adapted, effective codebook, which we denote as $\mathcal{C}_1$:
\begin{equation}
\mathcal{C}_1 = \mathbf{P}_{\text{sem}}(\mathbf{C}_{\text{sem}}),
\label{eq:1}
\end{equation}
where $\mathbf{P}_{\text{sem}}$ maps the source codebook into the encoder's latent space of dimension $D$. For each frame $\mathbf{h}_t$, the quantization index $i_t$ and embedding $\mathbf{e}_{1,t}$ are found via nearest-neighbor lookup in this adapted codebook:
\begin{equation}
i_t = \arg\min_{k} \| \mathbf{h}_t - \mathbf{c}_{1,k} \|^2_2, \quad \text{where } \mathbf{c}_{1,k} \in \mathcal{C}_1.
\label{eq:2}
\end{equation}
% As we will demonstrate in our ablation studies (Fig.~\ref{fig:semantic_ablation}), this global transformation is critical for ensuring full codebook utilization and achieving superior reconstruction performance.
This global transformation is designed to encourage full codebook utilization and enhance reconstruction quality, as further examined in our ablation studies (Fig.~\ref{fig:semantic_ablation}).

% \jh{revised, pls check}

\paragraph{Residual Activation Module} \label{residual_activation}
The semantic embedding $\mathbf{e}_{1,t}$ captures content but discards perceptually crucial acoustic details. We define this information, which includes vital paralinguistic attributes such as speaker timbre, prosodic rhythm, and speaking style, as the acoustic residual $\mathbf{r}_t$:
\begin{equation}
\mathbf{r}_t = \mathbf{h}_t - \mathbf{e}_{1,t}.
\label{eq:3}
\end{equation}
Reliably representing this residual is critical. Recent benchmarks~\cite{paralbench} and representation learning methods~\cite{sse} have underscored that these rich paralinguistic cues are not merely acoustic artifacts but are central to a vast range of downstream tasks, such as speaker verification and emotion recognition.

% To quantize this residual with maximum efficiency, our second quantizer module, $\mathbf{Q}_2$, employs a single-layer VQ empowered by SimVQ~\cite{simvq}. 
To efficiently quantize this residual, our second quantizer module, $\mathbf{Q}_2$, uses a single-layer vector quantizer enhanced by SimVQ~\cite{simvq}.
Instead of learning a residual codebook directly, SimVQ reparameterizes it as the product of a frozen, randomly initialized coefficient matrix $\mathbf{C}_{\text{coeff}} \in \mathbb{R}^{K_2 \times d}$ and a learnable linear ``latent basis'' $\mathbf{W}_{\text{basis}} \in \mathbb{R}^{d \times D}$:
\begin{equation}
\mathcal{C}_2 = \mathbf{C}_{\text{coeff}} \times \mathbf{W}_{\text{basis}}.
\label{eq:4}
\end{equation}
During training, only $\mathbf{W}_{\text{basis}}$ is updated. Gradients flow back to this shared basis, globally updating the entire effective residual codebook $\mathcal{C}_2$. This guarantees full codebook activation for the $K_2$=1024 entries. The quantized residual embedding $\mathbf{e}_{2,t}$ is then found via nearest-neighbor lookup in $\mathcal{C}_2$.
This design provides a highly efficient acoustic detail encoder at minimal architectural cost.

\subsection{Training Objective}
SACodec is trained end-to-end within a GAN framework. The generator $G$ (the codec itself) is optimized via a composite loss function designed to balance reconstruction fidelity, perceptual quality, and quantization stability, while a set of discriminators $\{D_k\}$ is trained to distinguish real from generated audio.

Our overall generator loss $\mathcal{L}_G$ is a weighted sum of several standard components widely used in high-fidelity speech synthesis~\cite{hifigan, dac}:
\begin{equation}
\mathcal{L}_G = \lambda_{\text{rec}}\mathcal{L}_{\text{rec}} + \lambda_{\text{adv}}\mathcal{L}_{\text{adv}} + \lambda_{\text{feat}}\mathcal{L}_{\text{feat}} + \lambda_{c1}\mathcal{L}_{\text{com},1} + \lambda_{c2}\mathcal{L}_{\text{com},2}.
\label{eq:total_loss}
\end{equation}
These components include: (1) a multi-scale mel-spectrogram reconstruction loss ($\mathcal{L}_{\text{rec}}$) for spectral accuracy; (2) an adversarial loss ($\mathcal{L}_{\text{adv}}$) based on a powerful ensemble of multi-period (MPD) and multi-band multi-scale STFT (MS-STFT) discriminators to enhance perceptual realism; (3) a feature-matching loss ($\mathcal{L}_{\text{feat}}$) to stabilize GAN training; and (4) two commitment losses ($\mathcal{L}_{\text{com},1}$ and $\mathcal{L}_{\text{com},2}$) to regularize the encoder's outputs for the semantic and residual modules, respectively. 
The weights for the reconstruction, adversarial, and feature-matching losses are set to $\lambda_{\text{rec}}=45.0$, $\lambda_{\text{adv}}=1.0$, and $\lambda_{\text{feat}}=1.0$ respectively, consistent with standard practices in WavTokenizer~\cite{wavtokenizer}. The commitment losses are weighted asymmetrically ($\lambda_{c1}=25.0, \lambda_{c2}=5.0$). A stronger weight on the semantic branch ($\lambda_{c1}$) is necessary to enforce the alignment of encoder features with the fixed, external mHuBERT space, while the dynamically learned residual branch requires a weaker regularization. 
% The mathematical formulation of each loss component is detailed in Appendix~B.
% \zix{1. in the figure, only two losses can be found. 2. how to choose these hyperparameters.}

\begin{table*}[!t]
  \centering
  \begin{threeparttable}
    \begin{tabular}{clrcccccccc}
      \toprule
      Dataset & Model & Params  & Token  & Codebook & $Q$
 & Bitrate & UTMOS$\uparrow$ & PESQ$\uparrow$ & STOI$\uparrow$ & F1$\uparrow$ \\
       &  &[M]  &Rate  & Size &  & [kbps] &  &  &  &  \\
      \midrule
      \multirow{9}{*}{\rotatebox{90}{LibriTTS Test-clean}} 
      & GT & - & - & - & - & - & 4.0562 & - & - & - \\
      & DAC & 74.71 & 75 & 1024 & 8 & 6 & 3.6905 & 3.5215 & .9546 & .9710 \\
      & Encodec & 14.85 & 75 & 1024 & 8 & 6 & 3.0399 & 2.7202 & .9391 & .9527 \\
      & SpeechTokenizer & 103.68 & 50 & 1024 & 8 & 4 & 3.8794 & 2.6121 & .9165 & .9495 \\
      & FACodec &374.49  & 80 & 1024 & 6 & 4.8 & 3.4454 & 2.2532 & .9127 & .9402 \\
      \cmidrule(lr){2-11}
      & DAC & 74.71 & 75 & 1024 & 2 & 1.5 & 1.9152 & 1.5300 & .8453 & .8957 \\
      & Encodec & 14.85 & 75 & 1024 & 2 & 1.5 & 1.5551 & 1.5398 & .8462 & .8496 \\
      & SpeechTokenizer & 103.68 & 50 & 1024 & 3 & 1.5 & 2.4121 & 1.2668 & .7853 & .8353 \\
      & WavTokenizer & 80.55 & 75 & 4096 & 1 & 0.9 & \underline{3.9687} & \underline{2.4687} & \underline{.9194} & \textbf{.9394} \\
      & \textit{SACodec(Ours)}  & 75.17 & 75 & 1000 / 1024 & 2 & 1.5 & \textbf{4.0373} & \textbf{2.6937} & \textbf{.9317} & \underline{.9381} \\
      \midrule
      \multirow{9}{*}{\rotatebox{90}{LibriTTS Test-other}} 
      & GT & - & - & - & - & - & 3.4831 & - & - & - \\
      & DAC & 74.71 & 75 & 1024 & 8 & 6 & 3.1338 & 3.3429 & .9402 & .9598 \\
      & Encodec & 14.85 & 75 & 1024 & 8 & 6 & 2.6568 & 2.6818 & .9241 & .9338 \\
      & SpeechTokenizer & 103.68 & 50 & 1024 & 8 & 4 & 3.2851 & 2.3269 & .8811 & .9205 \\
      & FACodec &374.49  & 80 & 1024 & 6 & 4.8 & 2.9302 & 2.0321 & .8832 & .9080 \\
      \cmidrule(lr){2-11}
      & DAC & 74.71 & 75 & 1024 & 2 & 1.5 & 1.7443 & 1.5039 & .8218 & .8636 \\
      & Encodec & 14.85 & 75 & 1024 & 2 & 1.5 & 1.5132 & 1.5753 & .8291 & .8228 \\
      & SpeechTokenizer & 103.68 & 50 & 1024 & 3 & 1.5 & 2.0104 & 1.2241 & .7780 & .7445 \\
      & WavTokenizer & 80.55 & 75 & 4096 & 1 & 0.9 & \underline{3.4315} & \underline{2.2705} & \textbf{.9173} & \underline{.8907} \\
      & \textit{SACodec(Ours)}  & 75.17 & 75 & 1000 / 1024 & 2 & 1.5 & \textbf{3.4786} & \textbf{2.4016} & \underline{.9040} & \textbf{.9273} \\
      \midrule
      \multirow{9}{*}{\rotatebox{90}{LJSpeech}} 
      & GT & - & - & - & - & - & 4.3794 & - & - & - \\
      & DAC & 74.71 & 75 & 1024 & 8 & 6 &4.0415  &3.4294  &.9567  &.9670  \\
      & Encodec & 14.85 & 75 & 1024 & 8 & 6 &3.2281  &2.6636  &.9442  &.9555  \\
      & SpeechTokenizer & 103.68 & 50 & 1024 & 8 & 4 &4.2371  &2.6411  &.9346  &.9453  \\
      & FACodec &374.49  & 80 & 1024 & 6 & 4.8 & 3.9760 & 2.3234 & .9220 & .9338 \\
      \cmidrule(lr){2-11}
      \cmidrule(lr){2-11}
      & DAC & 74.71 & 75 & 1024 & 2 & 1.5 &1.8169  &1.4307  &.8487  &.8904  \\
      & Encodec & 14.85 & 75 & 1024 & 2 & 1.5 &2.3900  &\underline{2.0195}  &\underline{.9058}  &\textbf{.9326}  \\
      & SpeechTokenizer & 103.68 & 50 & 1024 & 3 & 1.5 &3.5281  &1.6965  &.8790  &.9154  \\
      & WavTokenizer & 80.55 & 75 & 4096 & 1 & 0.9 &\underline{3.8755}  &1.9532  &.9007  &.9106  \\
      & \textit{SACodec(Ours)}  & 75.17 & 75 & 1000 / 1024 & 2 & 1.5 &\textbf{3.9912}  &\textbf{2.4249}  &\textbf{.9224} &\underline{.9302}  \\
      \bottomrule
    \end{tabular}
    \caption{Objective reconstruction results across multiple datasets. We evaluate performance on \textbf{in-domain} clean (\textit{LibriTTS-clean}) and noisy (\textit{LibriTTS-other}) conditions, as well as on \textbf{out-of-domain} data (\textit{LJSpeech}) to assess generalization. Best performance in the low-bitrate (1.5\,kbps) regime is highlighted in \textbf{bold} and the second best is \underline{underlined}.
    $Q$: \#quantizers. }
    \label{tab:reconstruction_objective}
  \end{threeparttable}
\end{table*}

\section{Experimental Setup}
% \subsection{Experimental Setup}
\paragraph{Datasets and Baselines.}
Our model is trained on the ~585-hour LibriTTS corpus~\cite{libritts}, using randomly cropped 1-second segments of 24\,kHz audio. We evaluate performance across three conditions: in-domain clean (\textit{LibriTTS test-clean}) and noisy (\textit{test-other}) splits to assess robustness, and on the out-of-domain dataset (\textit{LJSpeech})~\cite{ljspeech} to measure generalization. Our model is benchmarked against a comprehensive suite of SOTA codecs, including RVQ-based models (\textbf{Encodec}, \textbf{DAC}), semantic-distilled (\textbf{SpeechTokenizer}), disentanglement-focused (\textbf{FACodec}), and single-codebook (\textbf{WavTokenizer}) approaches. 
% For a fair comparison, these baselines are configured to operate in a bitrate of 1.5\,kbps regime.
For fairness, all baselines are set to operate at 1.5\,kbps whenever possible.

% paragraph
\paragraph{Implementation and Training.}
Our model, SACodec, is built in PyTorch. It employs an Encodec-style encoder with an LSTM and a modern ConvNeXt-Attention decoder inspired by WavTokenizer. The core asymmetric dual quantizer is configured with a 1000-entry fixed semantic codebook ($K_1$) and a 1024-entry SimVQ-activated residual codebook ($K_2$). The model is trained end-to-end using the AdamW optimizer. 
% Detailed hyperparameters and training procedures are provided in Appendix A.

\paragraph{Evaluation Metrics.}
Our evaluation is twofold. \textbf{Reconstruction Quality} is measured using objective metrics—UTMOS, PESQ, STOI, and V/UV F1—and complemented by subjective MUSHRA listening tests~\cite{mushra}. 
\textbf{Semantic Capability} is assessed using the comprehensive ARCH benchmark~\cite{arch}, which comprises multiple downstream classification tasks across speech, music, and audio domains. 
% The specific datasets used from each domain are detailed in Appendix D.
We evaluate this at two levels: (1) \textit{Compressed Domain}, which tests the intrinsic semantic richness of the raw tokens, and (2) \textit{Reconstruction Domain}, a novel evaluation we introduce to measure end-to-end semantic fidelity by analyzing the reconstructed waveform. This second dimension is crucial as it reveals whether semantic information survives the full generation pipeline.

\section{Results and Analysis}
% To validate the performance of SACodec, we conduct a comprehensive evaluation. After detailing our experimental setup, we present results across three key dimensions: (1) reconstruction quality against SOTA codecs at low bitrates, (2) a novel dual-domain analysis of semantic representation, and (3) a series of ablation studies verifying our core architectural components.
\subsection{Main Results}

\subsubsection{Acoustic Reconstruction Quality.}
As shown in Table~\ref{tab:reconstruction_objective}, SACodec establishes a new performance record in reconstruction quality for codecs operating at 1.5\,kbps, showing robust performance across diverse evaluation conditions.

On in-domain clean speech (\textit{LibriTTS test-clean}), our model's superiority at 1.5\,kbps is unequivocal. 
% It achieves a UTMOS of 4.0373, representing a dramatic leap in quality over other 1.5\,kbps codecs, such as Encodec at 1.5551 and DAC at 1.9152.
It achieves a UTMOS of 4.0373, over 2.5x higher than Encodec (1.5551) and 2x higher than DAC (1.9152) at 1.5 kbps.
Furthermore, it surpasses the aggressively compressed 0.9\,kbps WavTokenizer, confirming that our architecture effectively translates a modest increase in bitrate into substantial fidelity gains.

The model's robustness is particularly evident on the challenging, noisy \textit{LibriTTS test-other} set. Here, SACodec's UTMOS of 3.4786 is not only the highest among all low-bitrate competitors but is also nearly identical to the ground-truth audio's score of 3.483, indicating exceptional performance in reverberant conditions. This level of fidelity is remarkable, as it also exceeds that of higher-bitrate models, including the 6\,kbps DAC (3.1338), the 4\,kbps SpeechTokenizer (3.2851), and even the much larger FaCodec (2.930).

To assess generalization, we evaluated the models on the out-of-domain \textit{LJSpeech} dataset. SACodec performs strongly with a UTMOS of 3.9912, surpassing the single-codebook WavTokenizer and remaining competitive with higher-bitrate models like the 6\,kbps DAC (4.0415). This demonstrates that our architecture generalizes well beyond the training corpus to new acoustic domains.

\begin{figure}[!t]
\centering
\includegraphics[width=\columnwidth]{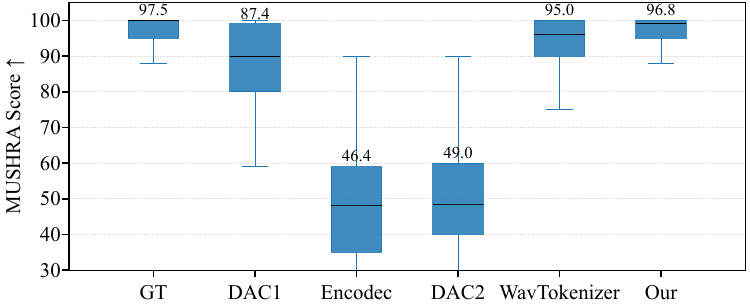} 
\caption{MUSHRA subjective evaluation on LibriTTS test-clean. Box plots show the distribution of listener scores (0-100, higher is better). The number above each box indicates the median score. DAC1: 6 kbps, DAC2: 1.5 kbps.}
\label{fig:mushra}
\end{figure}

\subsubsection{Subjective Evaluation}
To directly assess perceptual quality, we conducted a MUSHRA subjective listening test, with the results visualized as box plots in Fig.~\ref{fig:mushra}.The results affirm SACodec’s near-ground-truth quality.
At just 1.5\,kbps, SACodec achieves a high median score of 96.8, placing it in the same perceptual tier as the ground-truth audio's median of 97.5 and indicating consistently high quality. This stands in stark contrast to the low-bitrate DAC, which registered a low median of just 49.0 alongside a wide score distribution, signifying poor and inconsistent quality. Furthermore, SACodec maintains a clear performance edge over both the highly-regarded WavTokenizer and the 6\,kbps DAC baseline, the latter of which exhibits greater score variance while only reaching a median of 87.4. These results offer strong subjective evidence that our architecture produces speech with high naturalness and few audible artifacts.

\begin{table*}[!t]
  \centering
  \begin{threeparttable}
    \begin{tabular}{clcccccccccc} 
      \toprule
      \multirow{2}{*}{} & \multirow{2}{*}{Model} & \multirow{2}{*}{Token} & \multirow{2}{*}{$Q$} & \multirow{2}{*}{Bitrate } & \multicolumn{2}{c}{Speech$\uparrow$} & \multicolumn{2}{c}{Music$\uparrow$} & \multicolumn{2}{c}{Audio$\uparrow$} & \multirow{2}{*}{$Avg.$$\uparrow$} \\
      \cmidrule(lr){6-7} \cmidrule(lr){8-9} \cmidrule(lr){10-11}
      & &Rate & & [kbps] & RAVDESS & AM & MTT & MS-DB & ESC50 & VIVAE & \\
      \midrule
      % --- Compressed Domain Data ---
      \multirow{7}{*}{\rotatebox{90}{\textbf{Compressed}}}
      & Encodec & 75 & 8 & 6 & .3507 & .4913 & .3097 & .4226 & .2015 & .2675 & .3406 \\
      & DAC & 75 & 8 & 6 & .3889 & .7295 & .3331 & .5948 & .2440 & .3077 & .4330 \\
      & SpeechTokenizer & 50 & 8 & 4 & .4896 & .9725 & .3591 & .5566 & .3790 & .3027 & .5099 \\
      \cmidrule(lr){2-12}
      & Encodec & 75 & 2 & 1.5 & .2778 & .6135 & .2918 & .5001 & .2695 & .3079 & .3768 \\
      & DAC & 75 & 2 & 1.5 & \underline{.4236} & .6840 & \underline{.3023} & \textbf{.5859} & .2865 & .3166 & .4332 \\
      & FACodec  & 80 & 1 & 0.8 & .4231 & \underline{.8297} & .2766 & .5364 & \textbf{.3665} & \underline{.3311} & \underline{.4606} \\
      & WavTokenizer & 75 & 1 & 0.9 & .3438 & .6292 & .2689 & .5322 & .2340 & .2815 & .3816 \\
      & \textit{SACodec(Ours)}  & 75 & 1 & 0.75 & \textbf{.4265} & \textbf{.8845} & \textbf{.3281} & \underline{.5812} & \underline{.3385} & \textbf{.3331} & \textbf{.4809} \\
      \midrule
      
      % --- Reconstruction Domain Data ---
      \multirow{9}{*}{\rotatebox{90}{\textbf{Reconstruction}}}
      & Original & - & - & - & .8125 & .9985 & .4795 & .5978 & .6245 & .3910 & .6506 \\
      & DAC & 75 & 8 & 6 & .7743 & .9968 & .4751 & .5880 & .6285 & .4087 & .6452 \\
      & Encodec & 75 & 8 & 6 & .7778 & .9841 & .4762 & .5874 & .5945 & .3798 & .6333 \\
      & SpeechTokenizer & 50 & 8 & 4 & .7812 & .9940 & .4582 & .5790 & .5495 & .3609 & .6204 \\
      & FACodec  & 80 & 6 & 4.8 & .7083 &.9856 &.4766 &.5933 &.5615 &.3632 & .6148 \\
      \cmidrule(lr){2-12}
      & Encodec & 75 & 2 & 1.5 & .6667 & .9516 & \textbf{.4671} & .5852 & .5715 & .3656 & .6013 \\
      & DAC & 75 & 2 & 1.5 & .6667 & .9830 & \underline{.4641} & .5742 & \textbf{.5955} & .3418 & .6042 \\
      & SpeechTokenizer & 50 & 3 & 1.5 & .5729 & .9771 & .4364 & .5414 & .5150 & \underline{.3670} & .5683 \\
      & WavTokenizer & 75 & 1 & 0.9 & \underline{.6875} & \underline{.9906} & .4593 & \underline{.5948} & .5810 & .3398 & \underline{.6088} \\
      & \textit{SACodec(Ours)} & 75 & 2 & 1.5 & \textbf{.7569} & \textbf{.9933} & .4563 & \textbf{.5980} & \underline{.5825} & \textbf{.3997} & \textbf{.6311} \\
      
      \bottomrule
    \end{tabular}
    \caption{Semantic representation evaluation on the ARCH benchmark. We report classification accuracy for both the \textbf{Compressed Domain} and the \textbf{Reconstruction Domain}. Note that SACodec, WavTokenizer, and SpeechTokenizer were trained only on speech data, whereas DAC and Encodec utilized multi-domain (speech, music, audio) training data.}
    \label{tab:semantic}
  \end{threeparttable}
\end{table*}

% \subsection{Semantic Capability Evaluation}
% Beyond reconstruction, a crucial measure of a codec's utility is the semantic richness of its tokens. We assess this from dual perspectives.

\subsubsection{Semantic Representation Richness}
The results on the ARCH benchmark, summarized in Table~\ref{tab:semantic}, highlight the strong semantic capabilities of SACodec. Notably, SACodec and SpeechTokenizer were trained solely on speech data, while for fair comparison, we evaluate WavTokenizer using its speech-only trained model, despite its availability with multi-domain (speech, music, audio) training. In contrast, DAC and Encodec were trained on large-scale, multi-domain corpora.

In \textit{compressed domain}, this context makes SACodec's performance even more impressive. Its mean accuracy of 0.4809 considerably outperforms the speech-only WavTokenizer (0.3816) and, remarkably, remains highly competitive with the multi-domain trained DAC (0.4332), especially in the speech-related tasks. This validates that our semantic anchoring strategy effectively injects rich, generalizable semantic information.

In \textit{reconstruction domain}, SACodec demonstrates robust end-to-end semantic fidelity. Its mean score of 0.6311 not only surpasses the speech-only WavTokenizer but also achieves parity with the 6\,kbps DAC, which was trained on a much larger dataset. Crucially, SACodec's key advantage is its high consistency between the compressed and reconstruction domains.
In contrast, SpeechTokenizer—despite achieving the highest mean score in the compressed domain (0.5099)—fails to maintain this lead after low-bitrate reconstruction. This discrepancy suggests a potential pitfall in its decoder: an inability to fully preserve the intrinsic semantics of its tokens.
SACodec successfully avoids this pitfall. Its strong semantic fidelity even extends to unseen music and audio domains, underscoring the robustness of its asymmetric architecture.  While the multi-domain trained DAC maintains an edge on the music-specific MS-DB task, SACodec's superior performance on core speech-related tasks ultimately validates the effectiveness of our semantic anchoring strategy. 
% Notably, our performance advantage over the speech-only WavTokenizer is statistically significant in both the compressed and reconstruction domains ($\rho < .05$, one-tailed $t$-test).

\begin{table*}[!t]
  \centering
  \begin{threeparttable}
    \begin{tabular}{l cccc c}
      \toprule
      \multirow{2}{*}{\textbf{Model Configuration}} & \multicolumn{4}{c}{\textbf{Reconstruction Quality}} & \textbf{Semantic Acc. (C/R)} \\
      \cmidrule(lr){2-5} \cmidrule(lr){6-6}
      & UTMOS $\uparrow$ & PESQ $\uparrow$ & STOI $\uparrow$ & F1 $\uparrow$ & $Avg.\uparrow$ \\
      \midrule
      \textbf{SACodec (Full Model, $K_1$=1000, $K_2$=1024)} & \underline{4.0373} & \underline{2.6937} & \textbf{.9317} & \underline{.9381} & \textbf{.4809} / \underline{.6311} \\
       \midrule
      \textit{Ablation on Semantic Anchor ($\mathbf{Q}_1$)} \\
      \quad w/o Q1 (SimVQ-only) & 4.0132 & 2.6614 & .9301 & .9369 & .3494 / .6198 \\
      \quad w/ Random Learnable Codebook \tnote{a} & 3.9051 & 2.5967 & .9254 & .9337 & .3356 / .5966 \\
      \quad w/ Smaller Anchor ($K_1$=500) \tnote{b} & 3.8610 & 2.4981 & .9208 & .9307 & .4786 / .5823 \\
      \midrule
      \textit{Ablation on Residual Activator ($\mathbf{Q}_2$)} \\
      \quad w/o Q2 (Semantic-only) & 3.9461 & 2.3504 & .9058 & \textbf{.9407} & .5065 / .5765  \\
      \quad w/ Larger Residual ($K_2$=2048) & \textbf{4.0402} & \textbf{2.7010} & \underline{.9308} & .9379 & \underline{.4789} / \textbf{.6402} \\ 
      \bottomrule
    \end{tabular}
    \begin{tablenotes}[para]
      \small
      \item[a] Fixed mHuBERT codebook replaced with a standard learnable VQ.
      \item[b] Semantic anchor size ($K_1$) reduced to 500, using HuBERT-L9 k-means centroids. This size was chosen as it is the only other publicly available, comparable semantic codebook.
      % \item[e] Residual codebook size ($K_2$) increased from 1024 to 2048.
    \end{tablenotes}
  \caption{Ablation study of SACodec's core components and design choices. We evaluate reconstruction quality on LibriTTS test-clean and report mean semantic accuracy on the ARCH benchmark in the \textbf{C}ompressed / \textbf{R}econstruction domains.}
  \label{tab:ablation_study_comprehensive}
  \end{threeparttable}
  % \vspace{-.4cm}
\end{table*}

\begin{figure}[!t]
\centering
\includegraphics[width=\columnwidth]{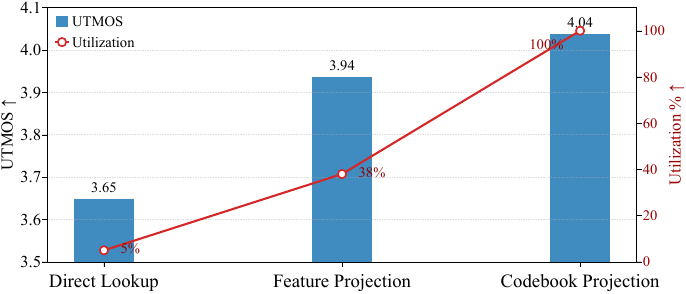} 
\caption{Comparison of semantic anchoring strategies. Our proposed Codebook Projection achieves near-perfect codebook utilization and significantly outperforms alternative strategies in reconstruction quality (UTMOS).}
% Full objective metrics are available in Appendix C.
\label{fig:semantic_ablation}
\end{figure}

\subsection{Ablation Study}
We conducted a series of targeted ablation studies to validate our asymmetric dual-quantizer architecture, with full results in Fig.~\ref{fig:semantic_ablation} and Table~\ref{tab:ablation_study_comprehensive}.

Our semantic anchoring strategy is critical for both utilization and quality. As shown in Fig.~\ref{fig:semantic_ablation}, a naive direct lookup between the encoder and the fixed semantic codebook results in a near-zero 5\% codebook utilization and a low UTMOS of 3.65. Our proposed codebook-space projection resolves this, achieving nearly 100\% utilization and boosting the UTMOS to 4.04, confirming its superiority over intermediate strategies like feature projection.

Each quantizer plays a distinct and indispensable role.
As shown in Table~\ref{tab:ablation_study_comprehensive}, the \textit{semantic anchor} ($\mathbf{Q}_1$) is unequivocally the source of semantic representation; removing it (`w/o Q1') or replacing its pre-trained knowledge with a standard learnable VQ (`w/ Random') causes semantic accuracy to collapse by up to 30\%. In parallel, the \textit{residual activator} ($\mathbf{Q}_2$) is essential for acoustic fidelity. Removing it (`w/o Q2') severely degrades reconstruction quality, causing a 12.8\% plunge in the PESQ score. Interestingly, this ``semantic-only'' configuration yields the highest semantic score (0.5065), highlighting an inherent tension between acoustic detail and semantic purity that our full model successfully reconciles through its synergistic design.
Furthermore, the residual codebook size ablation validates our design choice for bitrate efficiency. As Table~\ref{tab:ablation_study_comprehensive} shows, doubling the residual codebook size to 2048 (`w/ Larger Residual') offers only negligible gains in reconstruction quality (e.g., UTMOS increases from 4.0373 to 4.0402), confirming that our chosen size of $K_2$=1024 achieves a superior performance-to-bitrate trade-off.

\subsection{Discussion}
\paragraph{Training and Data Efficiency.}
SACodec exhibits significant training advantages. On the same 585h LibriTTS dataset, it achieves a \textgreater 6x training speedup per epoch over WavTokenizer due to its architectural aptitude for shorter audio chunks. Furthermore, it reaches SOTA performance on this public dataset, while baselines like Encodec, DAC and FACodec require massive proprietary corpora. All these highlight our model's superior data and computation efficiency.
% (see Appendix F for details)

\paragraph{Limitations and Future Work.}
Our study is confined to English; future work should evaluate cross-lingual robustness. While token-level evaluations indicate effective performance, direct integration into downstream SLMs and TTS systems is required for ultimate validation. Finally, scaling the semantic codebook and exploring model compression for on-device deployment are promising future directions~\cite{repvgg, hrf}.

\section{Conclusion}
This paper introduced Semantic-Anchored speech codec (SACodec), a neural speech codec that addresses the core trade-off between acoustic quality and semantic richness at low bitrates. At a mere 1.5 kbps, SACodec delivers reconstruction quality comparable to ground-truth audio while producing tokens with superior semantic expressiveness. This is made possible by our asymmetric dual-quantizer, a design that overcomes the limitations of traditional VQ by synergistically anchoring semantics and activating residuals. Ultimately, SACodec provides a blueprint for a new generation of codecs, demonstrating that fidelity, semantics, and architectural simplicity can be achieved in unison for modern Speech Language Models.

\section{Acknowledgments}
 This work was supported by the Beijing Xiaomi Mobile Software Co., Ltd, Beijing, China. In addition, the work leading to this research was supported by the National Natural Science Foundation of China under Grant No.~U25A20447 and No.~62571184, the Science and Technology Innovation Program of Hunan Province under Grant No.~2025RC6003, the Guangdong Basic and Applied Basic Research Foundation under Grant No.~2024A1515010112, the Changsha Science and Technology Bureau Foundation under Grant No.~kq2402082, and the Shenzhen Natural Science Foundation under Grant No.~JCYJ20250604190534043.

\bibliography{aaai2026}

\end{document}